\documentclass[runningheads]{llncs}
\usepackage{graphicx}
\usepackage{caption}
\usepackage{float}
\usepackage{array}
\usepackage{amsmath}
\usepackage[font=scriptsize,labelfont=bf]{caption}
\newcommand*\samethanks[1][\value{footnote}]{\footnotemark[#1]}
\newcolumntype{P}[1]{>{\centering\arraybackslash}p{#1}}
\begin{document}
\title{Memory Efficient 3D U-Net with \\ Reversible Mobile Inverted Bottlenecks for Brain Tumor Segmentation}
\titlerunning{3D U-Net with Reversible Mobile Inverted Bottlenecks}
%
\author{
    Mihir Pendse\thanks{equal contribution} \and
    Vithursan Thangarasa\samethanks \and
    Vitaliy Chiley \and
    Ryan Holmdahl \and
    \\ Joel Hestness \and
    Dennis DeCoste
    \\
    \texttt{mihirpendse89@gmail.com, \{vithu,joel\}@cerebras.net}
}
\authorrunning{Pendse, Thangarasa et al.}
%
\institute{Cerebras Systems, Los Altos CA 94022, USA}
\maketitle              
\begin{abstract}
\vspace{-10pt}
We propose combining memory saving techniques with traditional U-Net architectures to increase the complexity of the models on the Brain Tumor Segmentation (BraTS) challenge. The BraTS challenge consists of a 3D segmentation of a 240x240x155x4 input image into a set of tumor classes. Because of the large volume and need for 3D convolutional layers, this task is very memory intensive. To address this, prior approaches use smaller cropped images while constraining the model's depth and width. Our 3D U-Net uses a reversible version of the mobile inverted bottleneck block defined in MobileNetV2, MnasNet and the more recent EfficientNet architectures to save activation memory during training. Using reversible layers enables the model to recompute input activations given the outputs of that layer, saving memory by eliminating the need to store activations during the forward pass. The inverted residual bottleneck block uses lightweight depthwise separable convolutions to reduce computation by decomposing convolutions into a pointwise convolution and a depthwise convolution. Further, this block inverts traditional bottleneck blocks by placing an intermediate expansion layer between the input and output linear 1x1 convolution, reducing the total number of channels. Given a fixed memory budget, with these memory saving techniques, we are able to train image volumes up to 3x larger, models with 25\% more depth, or models with up to 2x the number of channels than a corresponding non-reversible network.

\end{abstract}
\section{Introduction}
Gliomas are a type of tumor affecting the glial cells that support the neurons in the central nervous system including the brain~\cite{GlioblastomaMuzaffar}. Gliomas are associated with hypoxia which causes them to invade and deprive healthy tissue of oxygen leading to necrosis. This can result in a range of symptoms including headaches, nausea, and vision loss. Brain gliomas are typically categorized into low grade glioma and high grade glioma based on their size and rate of growth with high grade gliomas having a much poorer prognosis and higher likelihood of recurrence after treatment. Diagnosing and treating gliomas early before they become serious is essential for improving the prognosis of the disease.

Magnetic resonance imaging (MRI) is one of the most commonly used imaging techniques used to identify neurological abnormalities including brain gliomas \cite{mrisurveyStefan}. One of the strengths of MRI is the ability to measure several different properties of tissue by adjusting the settings of the scan, namely the echo time and repetition time. For example, a scan with a short echo time and a short repetition time will result in a T1-weighted image that is sensitive to a property of tissues called spin-lattice relaxation which can help to differentiate between white and grey matter. A scan with longer echo and repetition times will result in a T2-weighted image that is sensitive to the spin-spin relaxation property of tissues and can be used to highlight the presence of fat and water. Another type of image called fluid attentuated inversion recovery (FLAIR) can be obtained by applying an inversion radiofrequency pulse that has the effect of nulling the signal from water making it easier to visualize lesions near the periphery of ventricles. Addition- ally, it is possible to inject a paramagnetic constrast agent such as gadolinium into the blood stream prior to the scan which will amplify signal from blood and make the vessels easier to visualize. Typically, for diagnosis of gliomas an MRI exam consists of T1-weighted, T1-weighted with gadolinium, T2-weighted, and FLAIR scans.

Based on an MRI exam, it is possible to identify four regions associated with the glioma \cite{Bakas2018IdentifyingTB}. At the center of the tumor is the region that is most affected and consists of a fluid filled necrotic core that is associated with high grade gliomas. The necrotic core is surrounded by a region called the enhancing region that hasn't undergone necrosis but still exhibits enhanced signal in T1-weighted images. Surrounding the enhancing region, is a region of the tumor that has reduced signal on T2-weighted images and is thus called the non-enhancing region. The core tumor consisting of the enhancing and non-enhancing regions is surrounded by peritumoral edematous tissue that is characterized by hyperintense signal on T2-weighted images and hypointense signal on T1-weighted images. Because the necrotic core is difficult to distinguish from the surrounding enhancing region the two can be grouped into the same class.\raggedbottom

Because of the heterogeneous nature of the composition and morphology of gliomas segmentation of these tumors on the MRI is time consuming even for experienced radiologists \cite{SIMI20151105}. For a human, the task of tracing an outline of various tumor classes on an imaging volume is limited by the two dimensional nature of human vision which requires iterating through several 2D slices in order view the entire volume. Furthermore, manual segmentation is subject variability between different radiologists and even between the same radiologist across multiple attempts. Computer vision algorithms could potentially help reduce the time needed for segmentation while also improving accuracy and reducing variance.

Convolutional neural networks have been shown to be a powerful class of machine learning models for extracting features from images to perform tasks such as classification, detection, and segmentation. For segmentation, the feature extraction stage of the network (encoder) is followed by a decoder that outputs a score for each output class. One of the first models proposed for segmentation called the Fully Convolutional Network (FCN) \cite{7298965} consists of 7 convolutional layers in the encoder each of which reduces the image size while increasing the number of features followed by a single deconvolutional layer consisting of either transposed convolution or bilinear interpolation for upsampling. Because the FCN does not consist of any fully connected layers, it can be used with images of any size. While the FCN had strong performance on segmentation of natural images, the drawback of the FCN architecture is that the encoder layers lose local information through several layers of filtering that cannot be recovered through the single decoder layer. The U-Net \cite{unetOlaf} was shown to perform better on medical image segmentation. Instead of a single layer in the decoder, the U-Net uses the same number of layers as in the encoder resulting in a symmetric U shaped architecture. The U-Net introduced skip connections between the output of each encoder layer and the input of the corresponding decoder layer. The advantage of the skip connections is that precise local information is retained and can be used by the decoder in achieving sharp segmentation outlines. The U-Net model is very popular in biomedical image segmentation due to its ability to segment images efficiently with a very limited amount of labeled training data. In addition, several variants of U-Net models have also been successfully implemented in various kinds of computer vision applications \cite{YAO2018364,iglovikov2018ternausnet,8575491}.

Although, U-Net models have been used successfully for many vision tasks, they are difficult to scale to high resolution images or 3D volumetric datasets. Activation memory requirements, which scale with network depth and mini-batch size, quickly become prohibitive. Thus, one of the main challenges with 3D segmentation of high-resolution MRIs is that the large volumetric images result in a high memory footprint to store the activations at the intermediate layers of the U-Net, which in effect limits the size of the network that can be used within the memory budget of modern deep learning accelerators. One approach for addressing this limitation is to crop the image volume into patches sampled at different scales to reduce activation memory \cite{kamnitsas2016deepmedic}. However, this strategy has limitations since it requires stitching together several cropped regions during inference which can be problematic at the border of these regions. Furthermore, cropping discards contextual information due to the lack of global context that can be used to increase the accuracy of the segmentation \cite{fb4b66446b3144f4b96c3c39ffc98cf0}. Another memory saving technique is to use multiple 2D slices and a less memory intensive 2D network but this also prevents full utilization of the entire context and can limit the power of the model.\raggedbottom

\section{Methods}
\subsection{Reversible Layers}
An alternative memory saving approach that does not compromise the expressive power of the model is to use reversible layers \cite{NIPS2017_6816,Thangarasa2019RevUp} that reduce the memory requirements in exchange for additional computation. If certain restrictions are imposed on a residual layer, namely that the input dimensions are identical to the output dimensions, it is possible to recover the input of that layer from the output. Therefore, the input activations do not need to be stored during the forward pass and can be reconstructed on-the-fly during backward pass to compute the gradients of the weights.

\begin{figure}[h]
\centering
\includegraphics[width=\textwidth]{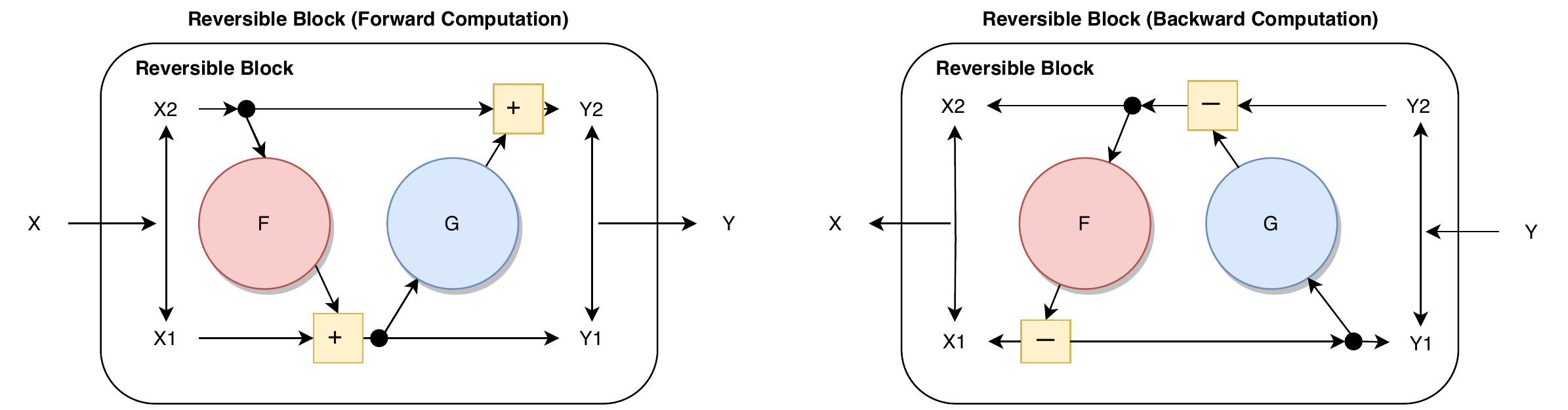}
\caption{The forward and backward computations of a reversible block.}
\label{fig0}
\end{figure}

The specific mechanism of a reversible layer is illustrated in Figure \ref{fig0}. During the forward computation, the input to the reversible layer is split across the channel dimension into two equally sized tensors $x_1$ and $x_2$. The $F$ and $G$ blocks represent two identical blocks (e.g., Convolution $\rightarrow$ Normalization $\rightarrow$ Non-linear Activation). The two output tensors $y_1$ and $y_2$ can be concatenated to get a tensor with same dimensions as the input. This can be expressed with the following equations:
\begin{equation}
  y_1 = x_1 + F(x_2) \quad \quad
  y_2 = x_2 + G(y_1).
  \label{eq:revf}
\end{equation}

On the backward pass, the input to the layer can be computed from the output as illustrated in Figure \ref{fig0}. The gradients of the weights of $F$ and $G$, as well as the reversible block’s original inputs are calculated. The design of the reversible block allows to reconstruct $x_1$ and $x_2$ given only $y_1$ and $y_2$ using Equation \ref{eq:revb}, thus making the block reversible.
\begin{equation}
  x_2 = y_2 - G(y_1) \quad \quad
  x_1 = y_1 - F(x_2).
  \label{eq:revb}
\end{equation}

It has been shown that for many tasks reversible layers maintain the same expressive power and achieve the same model accuracy as traditional layers with approximately same number of parameters. Reversible layers have been combined with the U-Net architecture to achieve memory savings by replacing a portion of the blocks in both the encoder and decoder with a reversible variant \cite{partUNet}.\raggedbottom

\begin{figure}[ht]
\centering
\includegraphics[width=\textwidth]{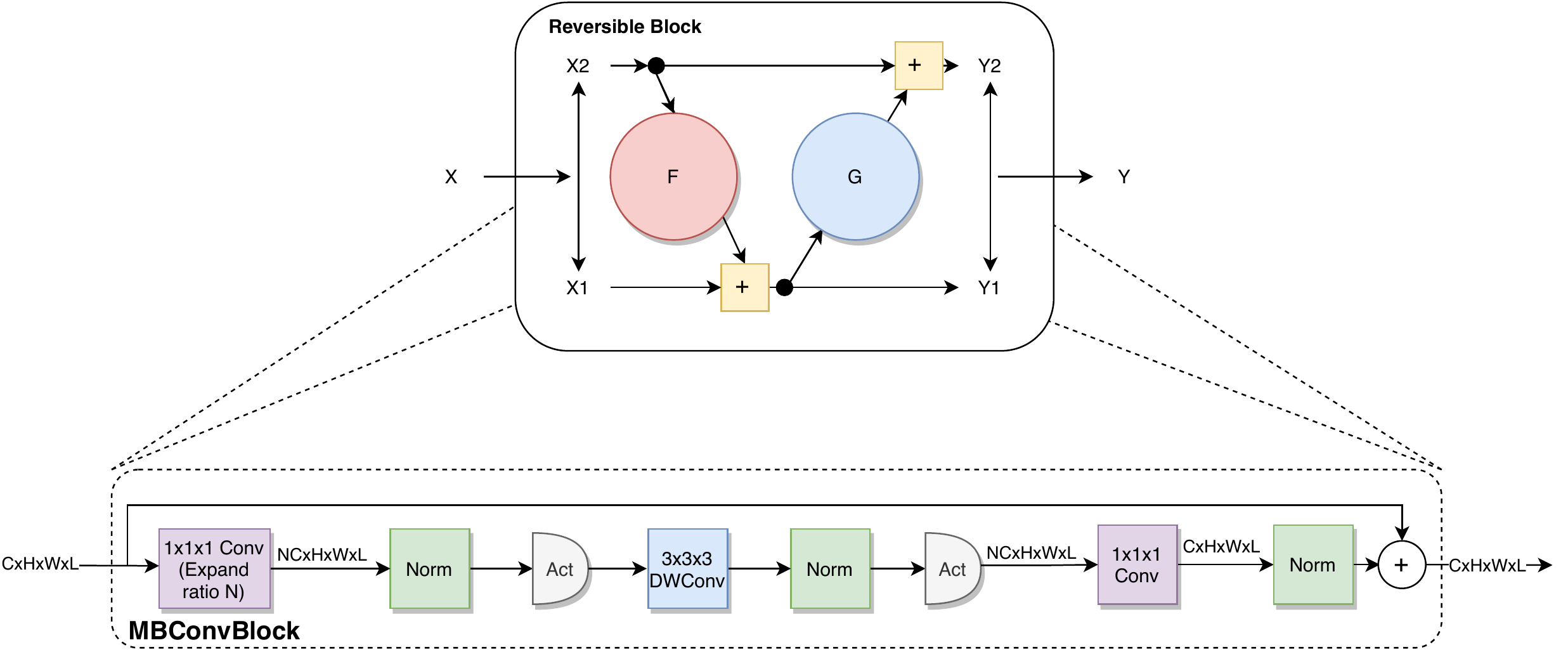}
\caption{The reversible MBConv block with inverted residual bottleneck and depthwise separable convolutions.}
\label{fig_block}
\end{figure}
\subsection{MobileNet Convolutional Block}
We introduce another memory saving technique that can be combined with reversibility to achieve additional performance by replacing traditional convolutional layers found in the standard U-Net with mobile inverted bottleneck convolutional block (MBConvBlock) introduced in MobileNetV2 \cite{DBLP:conf/cvpr/SandlerHZZC18} and later used in neural architecture search (NAS) based models such as MnasNet \cite{DBLP:conf/cvpr/TanCPVSHL19} and EfficientNet \cite{DBLP:conf/icml/TanL19}. The MBConvBlock consists of two important features. The components of this block are shown in Figure \ref{fig_block}. It replaces standard convolutions with depthwise separable convolutions consisting of a depthwise convolution (in which each input channel is convolved with a single convolutional kernel producing an output with same number of channels as the input) followed by a pointwise 1x1x1 convolution (where for each voxel a weighted sum of the input channels is computed to get the value of the corresponding voxel in the output channel). In the case of separable convolutions such as the Sobel filter for edge detection, it is possible to find values for the kernels of the depthwise and pointwise convolutions that make it mathematically identical to a standard convolution. More generally, even when the kernel of standard convolution is not separable, the loss in accuracy with a depthwise separable convolution is minimal and compensated for with reduction in total amount of computation \cite{DBLP:conf/icml/TanL19}. 

The second important feature of the MBConvBlock is the inverted residual with linear bottleneck block. In a conventional bottleneck block found in residual architectures such as ResNet-50 \cite{DBLP:conf/cvpr/HeZRS16}, the input to the block has a large number of channels and undergoes dimensionality reduction from convolutional layers with reduced number of channels before the final convolutional layer restores the original dimensionality. In an inverted residual block, the input has low dimensionality but the first convolutional layer consists of a pointwise convolution that results in expansion to a higher number of channels where the increase in dimensionality is given by a parameter called the expand ratio. This is followed by a depthwise separable convolution with the depthwise convolution occurring in the high dimensional space and the subsequent pointwise convolution projecting back into the lower dimensional space. This inverted bottleneck results in fewer number of parameters than a standard bottleneck block but also reduces the representational capacity of the network. To compensate for this, the nonlinear ReLU activation after the final convolutional layer is eliminated which was shown to improve accuracy in \cite{DBLP:conf/cvpr/SandlerHZZC18}.\raggedbottom

\subsection{Architecture}

\begin{figure}[h]
\centering
\includegraphics[width=\textwidth]{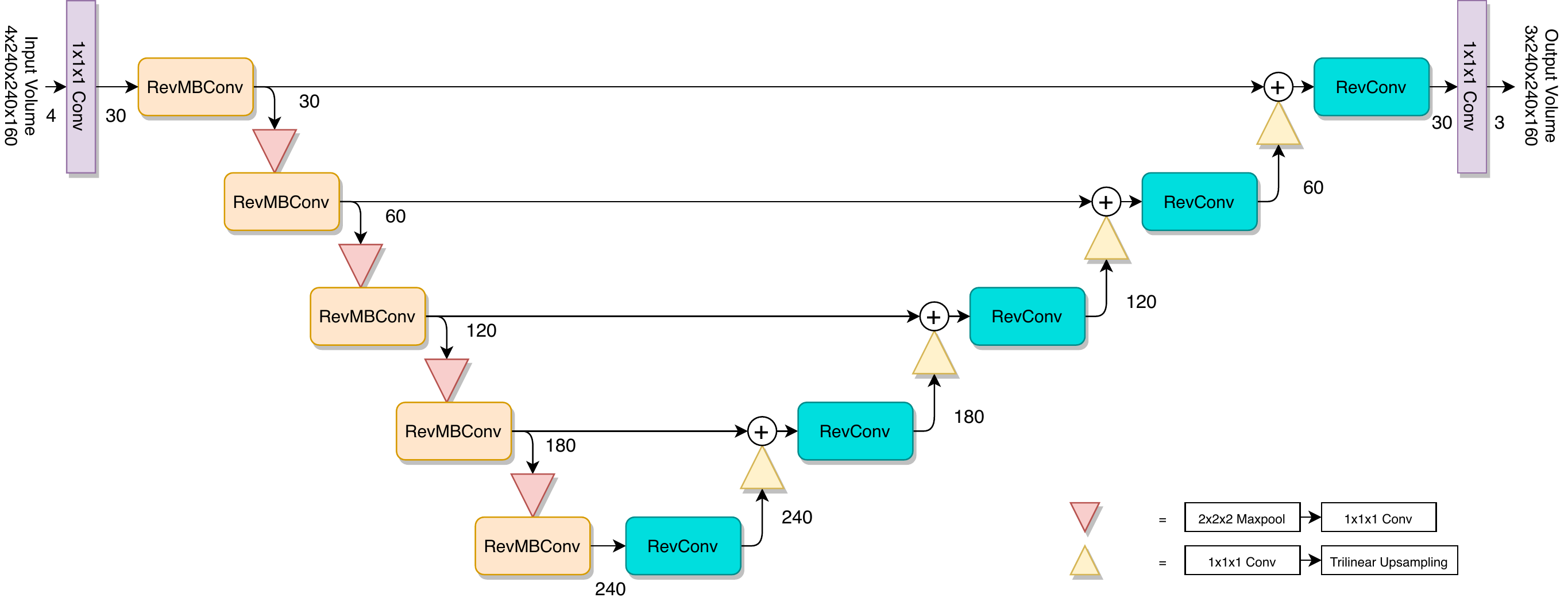}
\caption{Our reversible U-Net architecture with MBConv blocks in the encoder and regular convolutional blocks in the decoder. The downsampling and upsampling stages are depicted by red and yellow arrows, respectively. }
\label{fig_unet}
\end{figure}

Our architecture (Figure \ref{fig_unet}) consists of a U-Net with multiple levels of contraction in the encoder (through 2x2x2 max pooling) and the same number of levels of expansion in the decoder (through trilinear interpolation for upsampling instead of transposed convolutions as was shown to be preferable in \cite{partUNet}). Each level consists of two convolutional blocks. In the encoder, the first block is a pointwise convolution that increases the number of channels and the second block is a reversible block where each of the components (F and G in Figure \ref{fig0}) is a MBConvBlock with half the number of channels. We use additive instead of concatenated skip connections as in \cite{partUNet}. Because this memory intensive task requires using a batch size of 1, we use group normalization \cite{Wu_2018_ECCV} after the convolution instead of batch normalization.

\subsection{Training Procedure}
Training was done using Nvidia V100 GPUs for 500 epochs with initial learning rate of 0.0001 and learning rate drop by 5x at epoch 250 and 400. To speed up training, mixed precision and data parallel training with 4 GPUs (effective batch size of 4) was used resulting in a net speedup of about 5x compared to single GPU full precision training.

\noindent\textbf{Dataset:} The provided BraTS \cite{966861,brats2,Bakas2018IdentifyingTB,bratstcga,bratscga2} training dataset consists of 370 total examples each consisting of an MRI exam with 4 240x240x155 images (T1-weighted, Gadolinium enhanced T1-weighted, T2-weighted, and FLAIR) and a ground truth segmentation map grouping each voxel into one of four categories. We split this dataset into 330 examples for training and keep the remaining 40 examples as the hold out set for validation. \raggedbottom 

\noindent \textbf{Augmentation:} Because of the limited amount of data, we make extensive use of data augmentation to prevent overfitting. The augmentation applied includes the following: random rotation of the volume along the longitudinal axis by a random value between -20 and +20 degrees, random scaling up or down (resizing) of the image by at most 10\%, random flipping about each axis, randomly increasing or decreasing the intensity of the image by at most 10\%, and random elastic deformation.\raggedbottom

\section{Experiments and Results}

We compare four types of reversible U-Net architectures each with a constant 14GB of memory usage. The baseline consists of standard convolutional blocks for $F$ and $G$ in the reversible layers of the encoder. In the MBConv variants, $F$ and $G$ in the reversible layers of the encoder are replaced with the MBConv block. To make use of the additional memory, we explore using the full image volume (MBConv-Base), using a deeper model with cropped images (MBConv-Deeper), and a wider model with cropped images (MBConv-Wider).

\begin{table}[h]
\centering
\caption{Summary of experiments.}\label{tab1}
\begin{tabular}{|c|c|c|c|c|}
\hline
Experiment Name & Conv Block & Image Size & Channels & Expand ratio\\
\hline
Baseline & Standard & 256x256x160 & 60, 120, 180, 240, 480 & NA\\
MBConv-Base & MB & 256x256x160 & 30, 60, 120, 180, 240 & 2\\
MBConv-Deeper & MB & 128x128x128 & 30, 60, 120, 180, 240, 480 & 2\\
MBConv-Wider & MB & 128x128x128 & 30, 60, 120, 180, 240 & 8\\
\hline
\end{tabular}
\end{table}

\begin{figure}
\centering
\includegraphics[width=\textwidth]{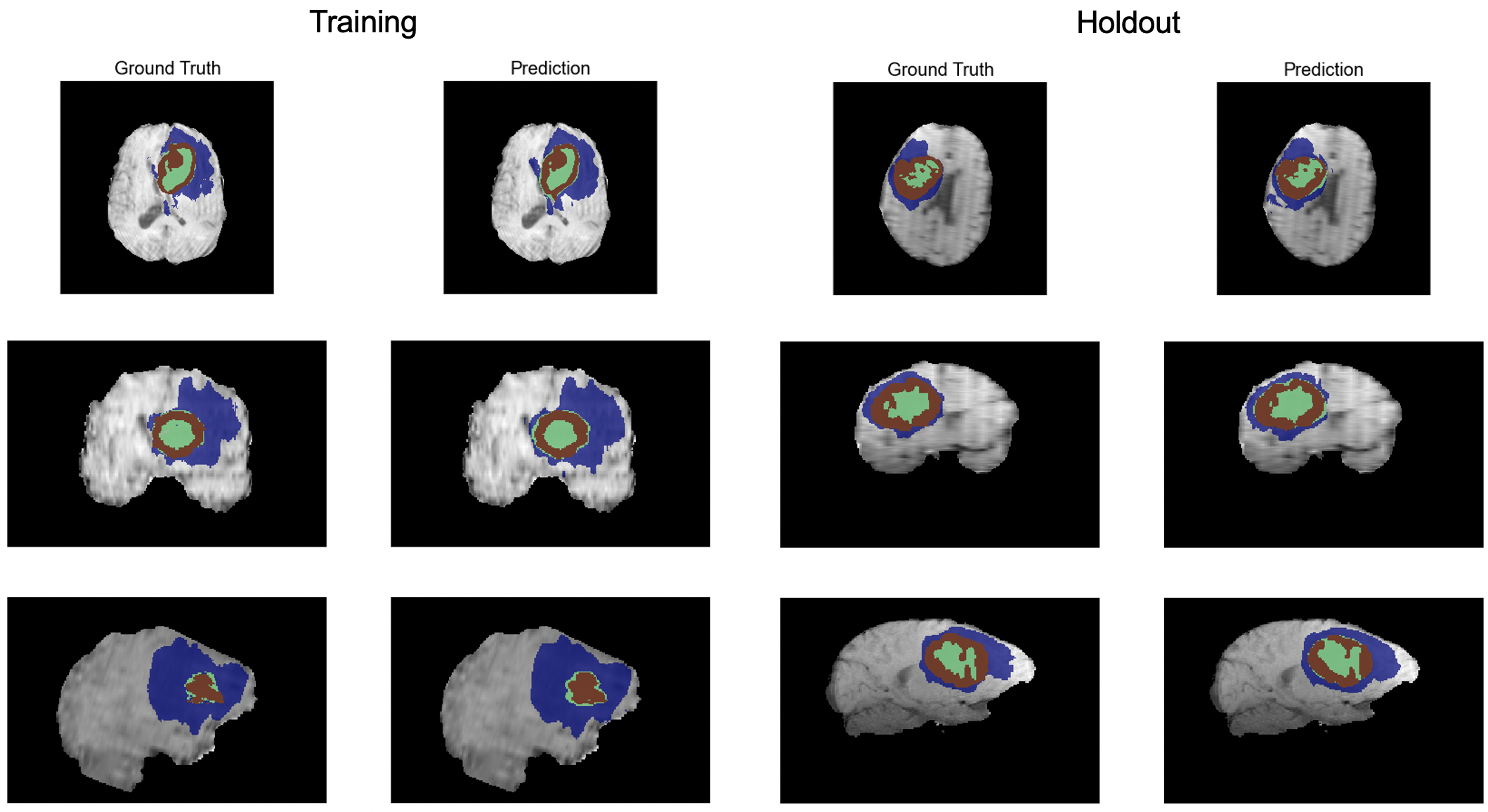}
\caption{Segmentation result for subject ID BraTS20\_Training\_210 (left) from training set (Dice\_ET = 0.88, Dice\_WT = 0.93, DICE\_TC = 0.92) and subject ID BraTS20\_Training\_360 (right) from holdout set (Dice\_ET = 0.92, Dice\_WT = 0.91, Dice\_TC = 0.95). Blue = whole tumor (WT), red = enhancing tumor (ET), green = tumor core (TC).}
\label{fig1}
\end{figure}

As seen in Table \ref{tab1}, our best MBConv reversible architecture was found to be the MBConv-Base variant which achieves a mean Dice score (averaged over all classes) above 0.7317 on hold out set after 50 epochs of training and Dice score of 0.7513 after convergence. The rate of convergence is faster than the baseline which only reaches a Dice score of 0.7184 after 50 epochs of training although the final score after convergence is slightly higher (0.7513). In Figure \ref{fig1}, a sample segmentation for an example from the the training set and an example from the holdout set indicate a close match between the prediction and the ground truth.

After identifying that the MBConv-Base variant performed the best, we trained three different models of this architecture to convergence using different initializations. We used following procedure to ensemble the three models to make the final prediction on the validation and test sets. For each image in the test set, a histogram of the pixel values was computed and the chisquared distance was computed with the histogram of each image in the training set. A weighted sum was computed across the training set for each model where the Dice score on each image was weighted by the chisquared distance of that image to the test image. The model with the lowest weighted sum was used to make the prediction for that particular test image

\begin{table}[h]
\centering
\caption{Experimental results.}\label{tab1}
\begin{tabular}{|c|c|c|c|}
\hline
Experiment Name & Dice Score after 50 epochs & Dice Score after convergence\\
\hline
Baseline & 0.7184 & 0.7513\\
MBConv-Base & 0.7317 & 0.7501 \\
MBConv-Deeper & 0.7129 & 0.7483\\
MBConv-Wider & 0.7092 & 0.7499\\
\hline
\end{tabular}
\end{table}

\raggedbottom

\section{Discussion}
We demonstrated the benefits of replacing a standard convolutional block with a MobileNet inverted residual with linear bottlneck block inside the reversible block of the encoder. This more parameter efficient MBConvBlock results in faster convergence while still fitting in a 16 GB GPU. For the same computational budget, the MBConvBlock gives more expressive power by replacing a single convolution with multiple convolutions in the form of a bottleneck block which has shown to improve accuracy on image classification tasks with architectures such as ResNet-50. When comparing the Dice score for an equal number of training steps, the MBConv-Basic variant is higher than the baseline. This is despite the fact that hyperparameters were tuned on the baseline model and the same values were used on the MB-Conv variant without further tuning. A significant drawback however is that the depthwise separable convolutions that are the dominant computation in the MB-Conv Block are slow on GPU. This is because standard convolutions are optimized to make use of the reuse of a convolutional kernel's weights on different inputs whereas in the depthwise separable convolutions does not have this optimization since each convolutional kernel is only applied to a single input. Therefore even though the MB-Conv block has fewer FLOPs than the standard one it is slower and results in longer wall clock time for each epoch. The fact that fewer epochs were needed for convergence suggests that the MB-Conv architecture is powerful and motivates optimizations to hardware that make depthwise separable convolutions efficient.

\bibliography{refs}

\begin{thebibliography}{10}
\providecommand{\url}[1]{\texttt{#1}}
\providecommand{\urlprefix}{URL }
\providecommand{\doi}[1]{https://doi.org/#1}

\bibitem{Bakas2018IdentifyingTB}
Bakas, S., Reyes, M., Jakab, A., Bauer, S., Rempfler, M., Crimi, A., Shinohara,
  R.T., Berger, C., Ha, S.M., Rozycki, M., Prastawa, M., Alberts, E.,
  Lipkov{\'a}, J., Freymann, J., Kirby, J., Bilello, M., Fathallah-Shaykh, H.,
  Wiest, R., Kirschke, J., Wiestler, B., Colen, R., Kotrotsou, A., LaMontagne,
  P., Marcus, D., Milchenko, M., Nazeri, A., Weber, M., Mahajan, A., Baid, U.,
  Kwon, D., Agarwal, M., Alam, M., Albiol, A., Varghese, A., Tuan, T., Arbel,
  T., Avery, A., Pranjal, B., Banerjee, S., Batchelder, T., Batmanghelich, N.,
  Battistella, E., Bendszus, M., Benson, E., Bernal, J., Biros, G., Cabezas,
  M., Chandra, S., Chang, Y.J., al., E.: Identifying the best machine learning
  algorithms for brain tumor segmentation, progression assessment, and overall
  survival prediction in the brats challenge. ArXiv  \textbf{abs/1811.02629}
  (2018)

\bibitem{brats2}
Bakas, S., Akbari, H., Sotiras, A., Bilello, M., Rozycki, M., Kirby, J.,
  Freymann, J., Farahani, K., Davatzikos, C.: Advancing the cancer genome atlas
  glioma mri collections with expert segmentation labels and radiomic features.
  Scientific Data  \textbf{4} (2017)

\bibitem{bratstcga}
Bakas, S., Akbari, H., Sotiras, A., Bilello, M., Rozycki, M., Kirby, J.,
  Freymann, J., Farahani, K., Davatzikos, C.: Segmentation labels and radiomic
  features for the pre-operative scans of the tcga-gbm collection. The Cancer
  Imaging Archive  (2017). \doi{10.7937/K9/TCIA.2017.KLXWJJ1Q}

\bibitem{bratscga2}
Bakas, S., Akbari, H., Sotiras, A., Bilello, M., Rozycki, M., Kirby, J.,
  Freymann, J., Farahani, K., Davatzikos, C.: Segmentation labels and radiomic
  features for the pre-operative scans of the tcga-lgg collection. The Cancer
  Imaging Archive  (2017). \doi{10.7937/K9/TCIA.2017.GJQ7R0EF}

\bibitem{mrisurveyStefan}
Bauer, S., Wiest, R., Nolte, L.P., Reyes, M.: A survey of mri-based medical
  image analysis for brain tumor studies. Physics in medicine and biology
  \textbf{58},  R97--R129 (06 2013)

\bibitem{partUNet}
Br{\"u}gger, R., Baumgartner, C.F., Konukoglu, E.: A partially reversible u-net
  for memory-efficient volumetric image segmentation. In: Shen, D., Liu, T.,
  Peters, T.M., Staib, L.H., Essert, C., Zhou, S., Yap, P.T., Khan, A. (eds.)
  Medical Image Computing and Computer Assisted Intervention -- MICCAI 2019.
  pp. 429--437. Springer International Publishing, Cham (2019)

\bibitem{NIPS2017_6816}
Gomez, A.N., Ren, M., Urtasun, R., Grosse, R.B.: The reversible residual
  network: Backpropagation without storing activations. In: Advances in Neural
  Information Processing Systems 30, pp. 2214--2224. Curran Associates, Inc.
  (2017)

\bibitem{GlioblastomaMuzaffar}
Hanif, F., Muzaffar, K., Perveen, k., Malhi, S., Simjee, S.: Glioblastoma
  multiforme: A review of its epidemiology and pathogenesis through clinical
  presentation and treatment. Asian Pacific Journal of Cancer Prevention
  \textbf{18}(1), ~3--9 (2017)

\bibitem{DBLP:conf/cvpr/HeZRS16}
He, K., Zhang, X., Ren, S., Sun, J.: Deep residual learning for image
  recognition. In: 2016 {IEEE} Conference on Computer Vision and Pattern
  Recognition, {CVPR}. pp. 770--778 (2016)

\bibitem{iglovikov2018ternausnet}
Iglovikov, V., Shvets, A.: Ternausnet: U-net with vgg11 encoder pre-trained on
  imagenet for image segmentation (2018)

\bibitem{fb4b66446b3144f4b96c3c39ffc98cf0}
Isensee, F., Kickingereder, P., Wick, W., Bendszus, M., Maier-Hein, K.: No
  new-net. In: Crimi, A., {van Walsum}, T., Bakas, S., Keyvan, F., Reyes, M.,
  Kuijf, H. (eds.) Brainlesion. pp. 234--244. Lecture Notes in Computer Science
  (including subseries Lecture Notes in Artificial Intelligence and Lecture
  Notes in Bioinformatics), Springer Verlag (Jan 2019), 4th International
  MICCAI Brainlesion Workshop, BrainLes 2018 held in conjunction with the
  Medical Image Computing for Computer Assisted Intervention Conference, MICCAI
  2018 ; Conference date: 16-09-2018 Through 20-09-2018

\bibitem{kamnitsas2016deepmedic}
Kamnitsas, K., Ferrante, E., Parisot, S., Ledig, C., Nori, A., Criminisi, A.,
  Rueckert, D., Glocker, B.: Deepmedic for brain tumor segmentation. In: MICCAI
  Brain Lesion Workshop (October 2016)

\bibitem{7298965}
{Long}, J., {Shelhamer}, E., {Darrell}, T.: Fully convolutional networks for
  semantic segmentation. In: 2015 IEEE Conference on Computer Vision and
  Pattern Recognition (CVPR). pp. 3431--3440 (2015)

\bibitem{966861}
Menze, B.H., Jakab, A., Bauer, S., Kalpathy-Cramer, J., Farahani, K., et~al.:
  The multimodal brain tumor image segmentation benchmark (brats). IEEE Trans
  Med Imaging  \textbf{34}(10),  1993--2024 (Oct 2015 2015).
  \doi{10.1109/TMI.2014.2377694}

\bibitem{unetOlaf}
Ronneberger, O., Fischer, P., Brox, T.: U-net: Convolutional networks for
  biomedical image segmentation. In: Navab, N., Hornegger, J., Wells, W.M.,
  Frangi, A.F. (eds.) Medical Image Computing and Computer-Assisted
  Intervention -- MICCAI 2015. pp. 234--241. Springer International Publishing,
  Cham (2015)

\bibitem{DBLP:conf/cvpr/SandlerHZZC18}
Sandler, M., Howard, A.G., Zhu, M., Zhmoginov, A., Chen, L.: Mobilenetv2:
  Inverted residuals and linear bottlenecks. In: 2018 {IEEE} Conference on
  Computer Vision and Pattern Recognition, {CVPR}. pp. 4510--4520 (2018)

\bibitem{SIMI20151105}
Simi, V., Joseph, J.: Segmentation of glioblastoma multiforme from mr images
  – a comprehensive review. The Egyptian Journal of Radiology and Nuclear
  Medicine  \textbf{46}(4),  1105 -- 1110 (2015)

\bibitem{8575491}
{Sun}, T., {Chen}, Z., {Yang}, W., {Wang}, Y.: Stacked u-nets with multi-output
  for road extraction. In: 2018 IEEE/CVF Conference on Computer Vision and
  Pattern Recognition Workshops (CVPRW). pp. 187--1874 (2018)

\bibitem{DBLP:conf/cvpr/TanCPVSHL19}
Tan, M., Chen, B., Pang, R., Vasudevan, V., Sandler, M., Howard, A., Le, Q.V.:
  Mnasnet: Platform-aware neural architecture search for mobile. In: 2019
  {IEEE} Conference on Computer Vision and Pattern Recognition, {CVPR}. pp.
  2820--2828. Computer Vision Foundation / {IEEE} (2019)

\bibitem{DBLP:conf/icml/TanL19}
Tan, M., Le, Q.V.: Efficientnet: Rethinking model scaling for convolutional
  neural networks. In: Proceedings of the 36th International Conference on
  Machine Learning, {ICML}. vol.~97, pp. 6105--6114 (2019)

\bibitem{Thangarasa2019RevUp}
Thangarasa, V., Tsai, C.Y., Taylor, G.W., K\"{o}ster, U.: Reversible fixup
  networks for memory-efficient training. In: NeurIPS Systems for ML (SysML)
  Workshop (2019)

\bibitem{Wu_2018_ECCV}
Wu, Y., He, K.: Group normalization. In: Proceedings of the European Conference
  on Computer Vision (ECCV) (2018)

\bibitem{YAO2018364}
Yao, W., Zeng, Z., Lian, C., Tang, H.: Pixel-wise regression using u-net and
  its application on pansharpening. Neurocomputing  \textbf{312},  364 -- 371
  (2018)

\end{thebibliography}
\bibliographystyle{splncs04}

\end{document}